\documentclass[12pt,floatfix]{revtex4}
\usepackage{graphicx}
\usepackage{amsmath}

\begin{document}

\title{Optical response of small magnesium clusters}

\author{Ilia A. Solov'yov}
\altaffiliation[Permanent address: ]{A. F. Ioffe Physical-Technical
Institute, Russian Academy of Sciences, Polytechnicheskaya 26,
194021 St. Petersburg, Russia }
\email[Email address: ]{ilia@th.physik.uni-frankfurt.de}
\author{Andrey V. Solov'yov}
\altaffiliation[Permanent address: ]{A. F. Ioffe Physical-Technical
Institute, Russian Academy of Sciences, Polytechnicheskaya 26,
194021 St. Petersburg, Russia}
\email[Email address: ]{solovyov@th.physik.uni-frankfurt.de}
\author{Walter Greiner}
\affiliation{Institut f\"{u}r Theoretische Physik, J.W. Goethe Universit\"{a}t,
Robert-Mayer str. 10, D-60054 Frankfurt am Main, Germany}

\begin{abstract}
We predict the strong enhancement in the photoabsorption of small $Mg$ clusters
in the region of 4-5 eV due to the resonant excitation of the plasmon oscillations of
cluster electrons.
The photoabsorption spectra for neutral $Mg$ clusters consisting of up to $N=11$
atoms have been calculated using {\it ab initio} framework based on the
time dependent density functional theory (TDDFT).
The nature of predicted resonances has been elucidated by comparison of the results of
the {\it ab initio} calculations with the results of the classical Mie theory.
The splitting of the plasmon resonances caused by the cluster deformation
is analysed. The reliability of the used calculation scheme has been proved by performing
the test calculation for a number of sodium clusters and the comparison of the 
results obtained with the results of other methods and experiment.

\end{abstract}
\pacs{36.40.-c, 36.40.Gk}
\maketitle

\section{Introduction}
Optical spectroscopy is a powerful instrument for investigation of the 
electronic and ionic structure of clusters as well as their thermal
and dynamical properties. During the last decades these issues have been
intensively investigated both experimentally by means
of photodepletion and photodetachment spectroscopy
and theoretically by
employing the time-dependent density functional theory
(TDDFT), configuration interaction (CI)
and random-phase approximation (RPA) (for review see \cite{Haberland94,MetCl99}
and references therein).
These methods have been used in conjunction with either jelium model \cite{MetCl99}
defined by a Hamiltonian, which treats the electrons in a cluster in the usual 
quantum mechanical way,
but approximates the field of the ionic core treating it as a uniform
positively charged background, or with {\it ab initio} calculations of the electronic
and ionic cluster structure, where all
or at least valence electrons in the system are
treated accurately.

During the last years, numerous theoretical and experimental investigations
have been devoted to the study of optical response properties
of alkali metal clusters. The plasmon resonances 
formation in Na, K and Li clusters has been
studied both theoretically and experimentally
(see \cite{Haberland94,MetCl99,DadCon,Klenig98,Landman01} and references therein).
Some attention was also devoted to the splitting and broadening
of the plasmon resonances (see citations above).
The mentioned metal elements belong to the first group
of the periodic table, i.e. possess one $s$-valence electron.

The situation differs for clusters of the alkali-earth metals of
the second group of the periodic table,
such as Be, Mg, Ca. Study of these clusters is of particular
interest, because they exhibit a transition from the
weak van der Waals bonding being the characteristic of the diatomic
molecule to the metallic bonding present in the bulk.
Thus, significant attention was paid to the 
magnesium clusters. Various properties of Mg clusters,
such as their structure, the binding energy, ionization
potentials, HOMO-LUMO gap, average distances, and their evolution with the cluster size
have been investigated
theoretically (see \cite{MgStruct,Jellinek02a,Jellinek02b} and references therein).
Recently, the mass spectrum of Mg clusters
was recorded \cite{Diederich01} and the sequence of magic numbers was determined.
The investigation of optical response of small Mg clusters has not been
performed so far in spite of the fact that it should carry a lot
of useful information about the dynamic properties of magnesium clusters.

In this paper we predict the strong enhancement in the photoabsorption of small
$Mg$ clusters in the region of 4-5 eV due to the resonant excitation of the plasmon
oscillations of the cluster electrons. Using all electron {\it ab initio} TDDFT
we calculate the spectra for cluster
structures with up to 11 atoms possessing the lowest energy. The geometries
of these clusters were calculated using all electron DFT methods and
described in our recent work \cite{MgStruct}. 
In this work we focus on the formation of the plasmon resonances
in magnesium clusters. We elucidate their nature by comparing our results
with the results of the classical Mie theory and analyse the splitting of the plasmon
resonances caused by the cluster deformation.

\section{Theoretical Method}

Theoretical methods used in our calculations are based on the density functional
theory and many-body-perturbation theory.
In the present work we use the 
gradient-corrected Becke-type three-parameter exchange functional \cite{Becke88}
paired with the gradient-corrected Lee, Yang and Parr correlation
functional (B3LYP) \cite{LYP}, as well as with the gradient-corrected
Perdew-Wang 91 correlation functional (B3PW91) \cite{PerWan}.
We do not  present the explicit forms
of these functionals, because
they are somewhat lengthy, and refer to the original papers
\cite{StructNa,Becke88,Gaussian98_man,LYP,PerWan}.
Our calculations have been 
performed  with the use of
the Gaussian 98 software package \cite{Gaussian98}.   
We have utilized the 6-311+G(d) basis set of 
primitive Gaussian functions to expand the cluster orbitals 
\cite{Gaussian98,Gaussian98_man}.

The absorption of light by small metal spheres has been investigated theoretically
by Mie long ago (see e.g. \cite{Vollmer}). For particles with the
diameter being considerably
smaller than the wavelength, the absorption cross section based on the Drude
dielectric function reads as:

\begin{equation}
\sigma(\omega)=\frac{4\pi N_ee^2}{m_ec}\frac{\omega^2\Gamma}{\left(\omega_0^2-\omega^2\right)^2+\omega^2\Gamma^2}
\label{PhotoAbsDrude}
\end{equation}

\noindent
where $\omega_0$ is the surface-plasma frequency of a sphere with $N_e$ free electrons,
$\omega$ is the photon frequency, $\Gamma$ represents the width of the resonance,
$m_e$ is the electron
mass, $e$ is its charge and $c$ is the light velocity.
Equation (\ref{PhotoAbsDrude}) assumes that the dipole oscillator strengths
are exhausted by the surface plasma resonance at $\omega_0$.
In metal clusters this resonance corresponds to the collective
oscillation of the spherical valence-electron cloud against the
positive background.

Using the sum rule one can easily show (see e.g. \cite{Vollmer}) that
$\omega_0=\sqrt{N_ee^2/m_e\alpha}$, where $\alpha$ is the static polarizability
of the cluster. For a classical metal sphere, $\alpha=N_er_s^3$, where
$r_s$ is the Wigner-Seitz radius. With $r_s=4.0$ a.u. for Na and $r_s=2.66$ for Mg
\cite{Kittel}, one derives the classical surface-plasma-resonance
energies $\omega_{0}^{Na}=3.40$ eV and
$\omega_{0}^{Mg}=6.27$ eV for $Na$ and $Mg$ respectively.

For small metal clusters the photoabsorption pattern differs significantly
from the Mie prediction.
In these systems the plasmon resonance energy is smaller as compared to the
metal sphere case.
The lowering of the plasmon energies in small metal clusters occurs
because of the spill out effect according to which
the electron density is spilled out of the cluster, increasing
its volume and polarizability.
For example, for spherical $Na_8$ and $Na_{20}$ clusters the
average static polarizability is 796.840 (a.u) and 1964.484 (a.u.)
respectively \cite{StructNa}. Thus, the plasmon resonance energies, $\omega_0$, read
as 2.73 and 2.75 (eV) for $Na_8$ and $Na_{20}$ respectively.
Beside the lowering of the plasmon resonance energy in small metal clusters
the photoabsorption pattern is splitted.
This fragmentation arises mainly due to the cluster deformation.
With the use of the sum rule, equation (\ref{PhotoAbsDrude})
can be generalized and written in the following form
(see e.g. \cite{Vollmer}):

\begin{equation}
\sigma(\omega)=\frac{4\pi e^2}{m_ec}\sum_{i=1}^n\frac{f_i\omega^2\Gamma_i}{\left(\omega_i^2-\omega^2\right)^2+\omega^2\Gamma_i^2}
\label{PhotoAbs}
\end{equation}

\noindent
where $\omega_i$ are the transition energies, $f_i$ and $\Gamma_i$ are the corresponding
oscillator strengths and widths, $n$ is the total number of the resonant transitions. 

In the case of the triaxial cluster deformation the photoabsorption cross section
possesses the three peak structure.
The splitting of the plasmon resonance into three peaks
can easily be understood assuming the ellipsoidal form of the cluster surface.
Within the framework of the deformed jelium model the ionic density is
considered to be uniform within the volume confined by the ellipsoid surface defined by
$\frac{x^2}{a^2}+\frac{y^2}{b^2}+\frac{z^2}{c^2}=1$.
If one assumes that the electron density fills in entirely in the interior of the
ionic ellipsoid, one finds the following dipole plasmon energies
corresponding to the electron density oscillations in
three directions $x$, $y$, $z$, (for more details see \cite{Lipparini}):

\begin{eqnarray}
\label{split}
\omega_x&=&\omega_{0}\left[1+\frac{\delta cos\gamma}{5}(1-\sqrt{3}tan\gamma)\right]\\
\nonumber
\omega_y&=&\omega_{0}\left[1+\frac{\delta cos\gamma}{5}(1+\sqrt{3}tan\gamma)\right]\\
\nonumber
\omega_z&=&\omega_{0}\left[1-\frac{2\delta cos\gamma}{5}\right]\\
\nonumber
\end{eqnarray}

\noindent
Where $\omega_{0}$ is the classical Mie frequency being the average
of $\omega_x$, $\omega_y$ and $\omega_z$, $\delta$ and $\gamma$ are 
the deformation parameters defined by equations:
$\delta cos\gamma=\frac{3}{4}\frac{2c^2-a^2-b^2}{a^2+b^2+c^2}$, 
$tan\gamma=\sqrt{3}\frac{a^2-b^2}{2c^2-a^2-b^2}$.
Note that in the axially symmetric case one derives $\gamma=0$ and $\omega_x=\omega_y$.

\section{Results and Discussion}

\begin{figure}
\begin{center}
\includegraphics[scale=0.57,clip]{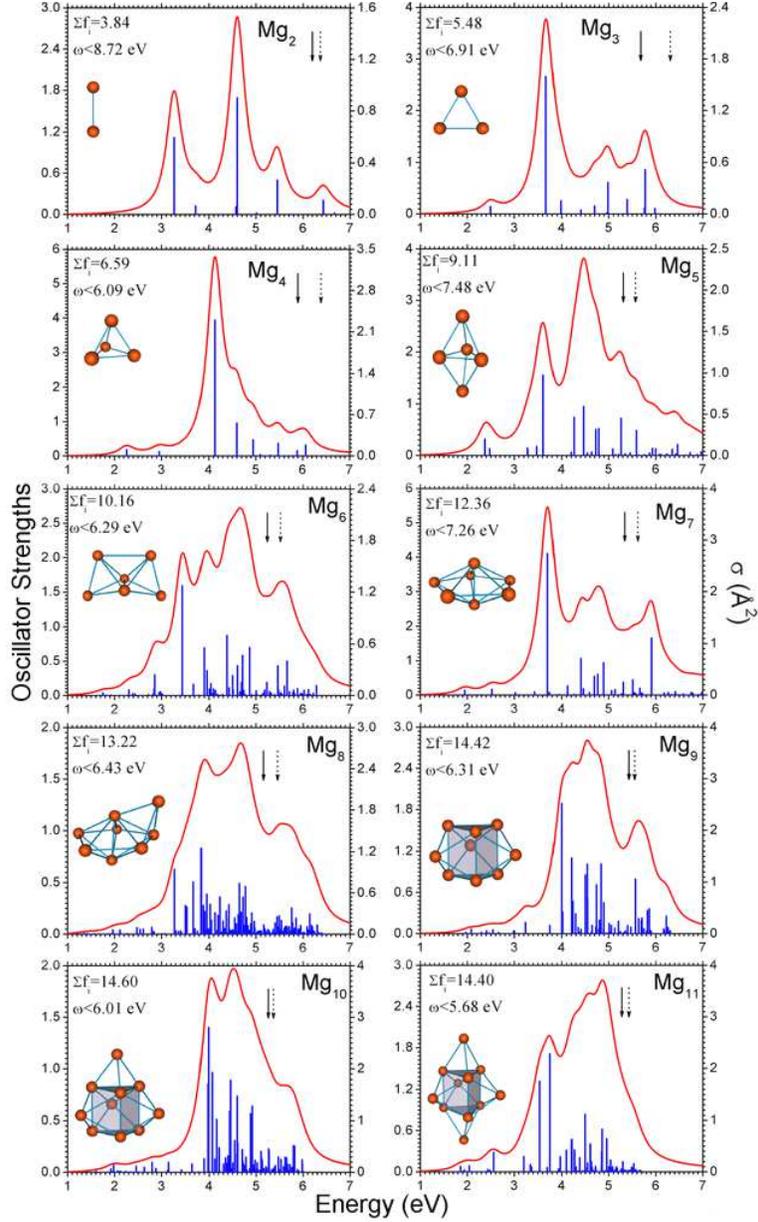}
\end{center}
\caption{
Photoabsorption cross section calculated for $Mg$ clusters
with $N\leq 11$ using the $B3PW91/6-311+G(d)$ method. Vertical solid
lines show the oscillator strengths for the optically allowed transitions.
Their values are shown in the left hand side of the plots. The right hand side
of each plot shows the scale for the corresponding photoabsorption
cross section.
Cluster geometries calculated in \cite{MgStruct} are shown in the insets.
The label near each cluster image shows the sum of the
oscillator strengths and the excitation energy range considered. By solid and dotted
arrows we show the adiabatic and vertical ionization potentials respectively,
calculated in \cite{MgStruct}.}
\label{spectra_Mg}
\end{figure}

In figure \ref{spectra_Mg},
we present the oscillator strengths for the dipole
transitions calculated for the most stable cluster isomers of $Mg_2$-$Mg_{11}$.
Cluster geometries are shown in the insets to the figure.
They were calculated and discussed in \cite{MgStruct}.

For sodium \cite{MetCl99}, the plasmon resonance
arises for the clusters with less than 10 atoms.
Thus, it is natural to expect that for the magnesium clusters with $N\leq 10$
the formation of the plasmon resonance should be clearly seen.

Calculating the oscillator strengths $f_i$ and substituting the found values
in equation \ref{PhotoAbs}, we obtain the photoabsorption cross
sections for magnesium clusters plotted in figure \ref{spectra_Mg}.
In this calculation we have used the width $\Gamma_0=0.4$ eV, 
which is the average width for $Na$ clusters at room temperature \cite{MetCl99}.
In this paper we do not 
calculate the excitation line widths for $Mg$ clusters and do not investigate
the line widths temperature dependence.
These interesting problems are beyond the scope of the present paper and
deserve a separate careful consideration.

\begin{figure}
\begin{center}
\includegraphics[scale=0.75,clip]{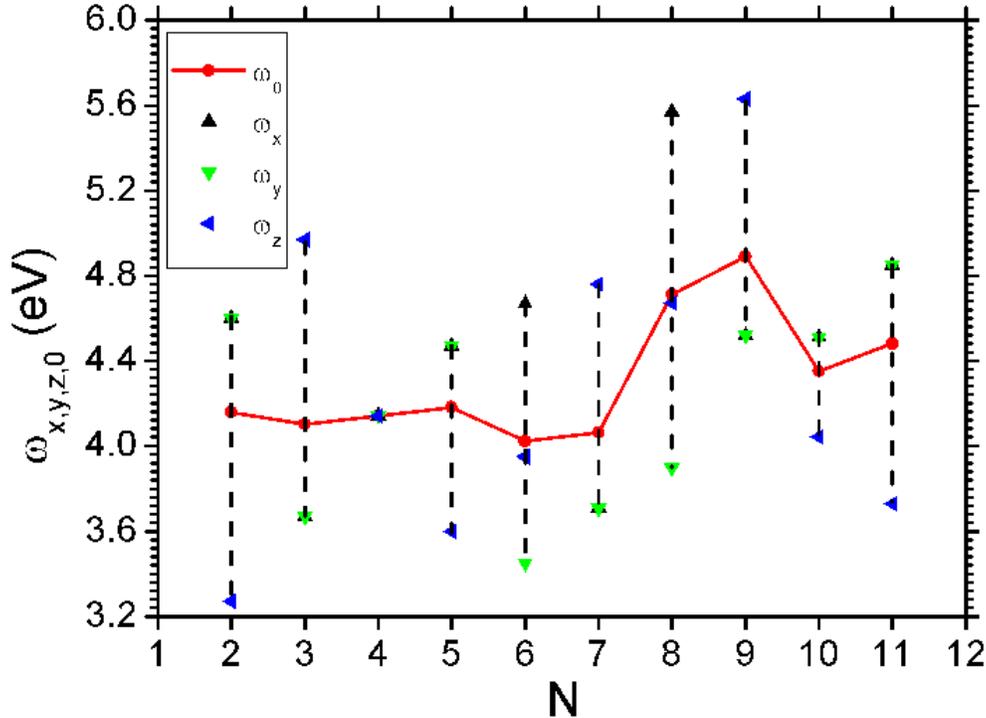}
\end{center}
\caption{
Size dependence of the plasmon
resonance energies $\omega_x$, $\omega_y$, $\omega_z$:
x (upper triangles), y (lower triangles) and z (left triangle).
Circles are the Mie-frequencies $\omega_0$ being the average of $\omega_x$, $\omega_y$
and $\omega_z$.}
\label{energies}
\end{figure}

In the photoabsorption spectra
for $Mg_2$ and $Mg_3$ one can identify the strong resonances in the vicinity of 4 eV,
which can be interpreted as the plasmon resonances splitted due to the cluster deformation.
Below, we discuss this splitting in more detail.
For larger clusters, the plasmon resonance energy increases slowly and evolves towards
the bulk value, 6.26 eV, see dots in figure \ref{energies}.
The lowering of the plasmon resonance energy in small Mg clusters as compared to its
bulk value occurs because of the spill out effect.

There are two main factors, which determine the resonance
pattern of the photoabsorption
spectra for magnesium clusters: collective plasmon excitations of
the delocalized electrons and
the resonant transitions of the electrons bound in a single magnesium atom.
In the excitation energy range considered, the photoabsorption
spectrum of a single Mg atom exhibits the two strong resonant excitations:
$3s(^1S_0)\rightarrow 3p(^1P_1^0)$ and $3s(^1S_0)\rightarrow 4p(^1P_1^0)$
with the  energies (oscillation strengths) 4.346 (1.8) 
and 6.118 (0.2) eV respectively \cite{RS}. The TD/B3PW91/6-311+G(d) method
gives the following  energies and the oscillator strengths for these lines:
4.225 (1.63) and 5.765 (0.29) eV,
which are in the reasonable agreement with the data given in \cite{RS}.
The $3p(^1P_1^0)$ line can be easily identified in the photoabsorption spectrum for $Mg_2$. 
In terms of the plasmon resonance excitations,
this line  corresponds to the oscillations of the electronic density perpendicular to
to the cluster axis, while the strong line in the vicinity of 3 eV
corresponds to the collective electron oscillations along the cluster axis.
For larger clusters, the $3p(^1P_1^0)$
line is strongly coupled with the plasmon resonance excitation occurring at the close
energy.
The situation is different for the $4p(^1P_1^0)$ line. Due to its higher
energy, this excitation line does not couple that strongly with the plasmon resonance
and can be identified in the photoabsorption spectra for
the $Mg_2$, $Mg_3$, $Mg_6$ and $Mg_7$ clusters 
in addition to the plasmon resonances.
For larger clusters (e.g. $Mg_8$, $Mg_9$, $Mg_{10}$), due to the growth 
of their plasmon resonance energies,
the $4p(^1P_1^0)$ line becomes more and more of the plasmon resonance type.

\begin{figure}
\begin{center}
\includegraphics[scale=0.75,clip]{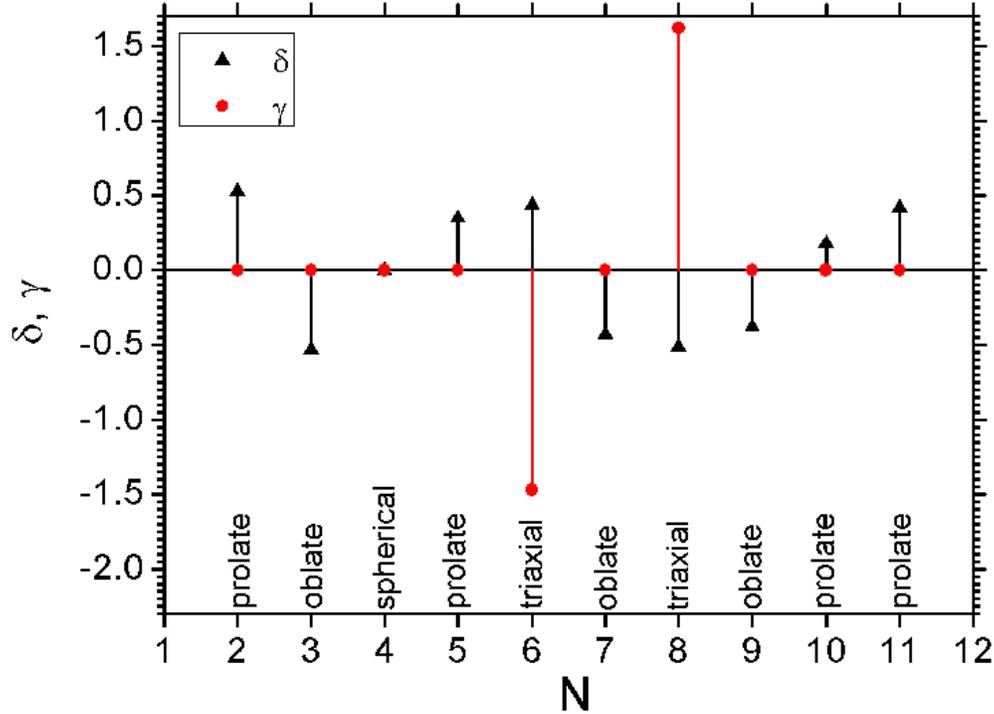}
\end{center}
\caption{
Cluster deformation parameters versus the
cluster size. The labels indicate the cluster deformation type.}
\label{deformation}
\end{figure}

For many clusters the plasmon resonance
is splitted. This splitting arises mainly due to the cluster
deformation. In order to illustrate this effect we plot in figure \ref{energies}
the energies $\omega_x$, $\omega_y$, $\omega_z$ of the strongest resonances
versus the cluster size. Using equation (\ref{split}),
we determine the deformation parameters $\delta$ and
$\gamma$ and present them in figure \ref{deformation}.
One can distinguish four different cases:
i)  $\delta=\gamma=0$ the cluster is spherical (see $N=4$);
ii) $\delta<0$, $\gamma=0$ the cluster is oblate (see $N=3,7,9$);
iii) $\delta>0$, $\gamma=0$ the cluster is prolate (see $N=2,5,10,11$);
iv) $\delta\ne 0$, $\gamma\ne 0$ the cluster is triaxially deformed
(see $N=6,8$).
This analysis shows that most of the clusters considered are close to the
axially symmetric form, although some clusters ($Mg_6$ and $Mg_8$)
are triaxially deformed.
Note that many additional satellite resonances appear in the photoabsorption
spectra. The additional satellite lines are often the result of higher order
cluster deformations. Thus, they are beyond the ellipsoidal model.

To show the connection between the plasmon resonance splitting and the cluster deformation
we have determined the plasmon resonance energies for $Mg_2$ and $Mg_3$ from
the Mie theory via
the static dipole polarizabilities of the clusters
and compared them with the TDDFT result. The
principle values of cluster polarizability tensor $\alpha_{xx}$, $\alpha_{yy}$,$\alpha_{zz}$
are 130.386, 130.386, 246.769 (a.u) for $Mg_2$ and
282.412, 282.412, 159.757 (a.u.) for $Mg_3$ respectively.
Thus, the plasmon resonance energies $\omega_x$, $\omega_y$ and $\omega_z$
read as 4.82, 4.82, 3.39 (eV) for $Mg_2$ and 3.8, 3.8, 5.46 (eV)
for $Mg_3$ respectively. These values are very close to those obtained directly from
the photoabsorption spectra analysis
and presented in figure \ref{energies}. This fact independently
proves that the plasmon resonance is already formed in such small systems.

In insets to figure \ref{spectra_Mg},
we present the sum of the oscillator strengths
and the excitation energy range considered for each cluster. The sum of the oscillator
strengths characterises the valence
electrons delocalization rate. Note, that for many clusters it is close
to the total number of valence electrons in the system. For some
clusters the total sum of the oscillator strengths is significantly smaller
than the number of the valence electrons (see, for example, $Mg_{10}$, $Mg_{11}$).
To increase the sum of the oscillator strengths
one has to calculate the photoabsorption spectra up to the higher excitation energies.
The calculation of cluster excited states becomes an increasingly difficult
problem with the growth of the cluster size, because of the rapid growth of
the number of possible excited states in the system.
In this paper we focus on the investigation of the plasmon resonances in small $Mg$
clusters, manifesting themselves in the energy range about 4-5 eV as
it is clear from our discussion. Therefore, for clusters with $N>8$,
we have calculated the photoabsorption spectra only up to the
excitation energies of about 6 eV, which is significant for the elucidation of the
plasmon resonance structure and at the same time it does not acquire substantial
computer power.

\begin{figure}
\begin{center}
\includegraphics[scale=0.65,clip]{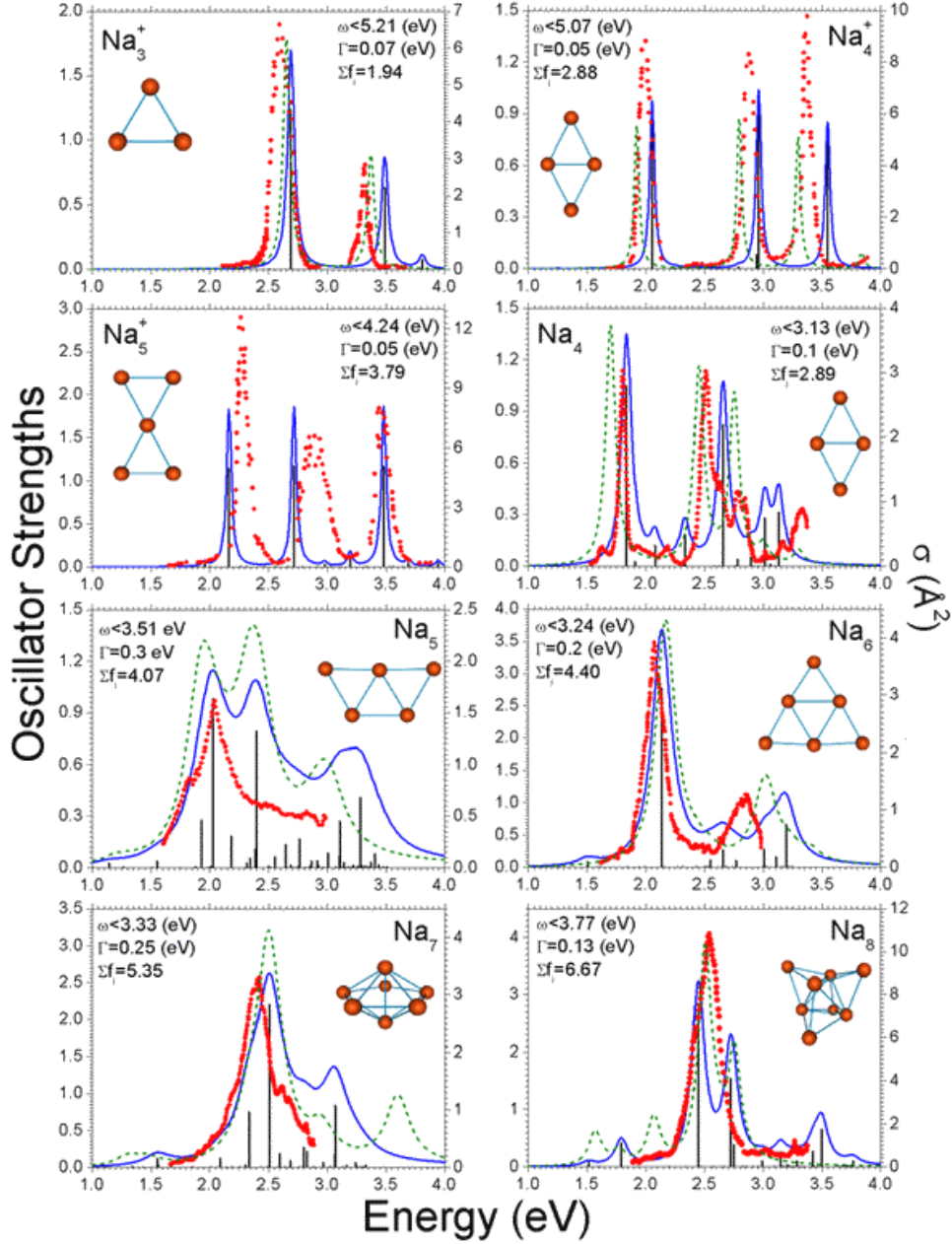}
\end{center}
\caption{
Photoabsorption cross section calculated for $Na_{3-5}^+$,
$Na_{4-8}$ using the B3LYP functional (solid lines).
Vertical lines show the oscillator strengths for the optically allowed transitions.
Cluster geometries calculated in \cite{StructNa} are shown in the insets.
The label near each cluster image shows the sum of the
oscillator strengths, the excitation energy range considered and the line width.
We compare our results with experimentally measured photoabsorption
spectra \cite{Haberland94, MetCl99} (dots)
and with the results of previous {\it ab initio} CI calculation
\cite{Haberland94, MetCl99} (dashed lines).}
\label{spectra_Na}
\end{figure}

Photoabsorption spectra for sodium clusters have been
earlier investigated in a large number of papers. There were performed
experimental measurements,
as well as theoretical calculations \cite{Haberland94, MetCl99} involving
{\it ab initio} and model approaches. 
In order to check the level of accuracy of our calculation method,
in figure \ref{spectra_Na}, we compare  the photoabsorption
spectra for a few selected neutral and singly charged sodium clusters, calculated
with the use of the methods
described above, with the results of experimental measurements and
other calculations. In figure \ref{spectra_Na}, the
experimentally measured photoabsorption
spectra for $Na_{3-5}^+$, $Na_{4-8}$
are plotted by dots.
The results of our TDDFT calculation performed with the use of the
B3LYP functional are shown by solid lines. The CI results
of Bona\v{c}i\'{c}-Kouteck\'{y} {\it et al} \cite{Haberland94, MetCl99} 
are shown by dashed lines.

In \cite{StructNa} we demonstrated that the B3LYP functional is well applicable for
the description of sodium clusters.
Thus, we used it for the photoabsorption spectra computations.
The comparison shown in figure \ref{spectra_Na} demonstrates
that our calculation method is a good alternative to
the CI method, and our results are
in a good agreement with the experimental data.

The photoabsorption spectrum of $Na_5$ has a prominent peak at the energy about
2.3 eV, which can be identified as a Mie plasmon resonance.
This peak is also seen in the 
photoabsorption spectra of $Na_6$, $Na_7$ and $Na_8$.
The plasmon resonance energy for these clusters is smaller than the bulk value, 3.4 eV,
because of the spill out effect. As it
is seen from figure \ref{spectra_Na}, the resonance energy
evolves slowly towards the bulk limit with increasing cluster size.
For the $Na_3^+$, $Na_4^+$ and $Na_5^+$,
the plasmon peak is hardly to identify in the distribution of
oscillator strengths, which means that the number of delocalized electrons in these clusters
turns out to be insufficient for the formation of the plasmon resonance in this system,
see figure \ref{spectra_Na}.

Note, that often the plasmon peaks for sodium clusters
are splitted due to the cluster axial quadrupole deformation. 
Using equations (\ref{deformation}), we have calculated
the deformation parameters for axially symmetric $Na_6$ and $Na_7$.
The result reads as $\delta=-0.55$ and $-0.34$ respectively.
The deformation parameter $\gamma$ vanishes for both clusters.
The axially symmetric jelium model leads to the following
values of $\delta$ \cite{JM}: $\delta_{JM}=-0.48$ and $-0.24$
for $Na_6$ and $Na_7$ respectively.
Comparison shows that the splitting of the plasmon resonances can be explained
by cluster deformation.

\section{Conclusion}

In this paper we predict the enhancement of the photoabsorption spectra for small
Mg clusters in the vicinity of plasmon resonance.
The photoabsorption spectra for neutral $Mg$ clusters consisting of up to $N=11$
atoms have been calculated using {\it ab initio} framework based on the
time dependent density functional theory.
The nature of predicted resonances have been elucidated by comparison of the results of
the {\it ab initio} calculations with the results of the classical Mie theory.
The splitting of the plasmon resonances caused by the cluster deformation
is analysed. The reliability of the used calculation scheme has been proved by performing
the test calculation for a number of sodium clusters and the comparison of the 
results obtained with the results of other methods and experiment.
The calculation of the photoabsorption spectra for larger clusters
requires much more computer power and is left open for further investigations.

\section*{Acknowledgements}
The authors acknowledge support from the Russian Foundation for
Basic Research (grant No 03-02-16415-a), Russian Academy of Sciences
(grant No 44) and the Studienstiftung des deutschen Volkes.

\end{document}